\documentclass[rnote,referee]{aa}  
\usepackage{natbib}
\bibpunct{(}{)}{;}{a}{}{,} 
\usepackage{graphicx}
\usepackage{txfonts}
\newcommand{\va}{v_{\mathrm{A}}}
\newcommand{\cs}{c_{\mathrm{s}}}

\newcommand{\mutilde}{\tilde{\mu}}

\newcommand{\td}{\tau_{\mathrm{D}}}
\newcommand{\tdp}{\tau_{\mathrm{D}} / P}

\begin{document}

	\title{Time damping of non-adiabatic magnetohydrodynamic waves in a partially ionized prominence plasma: Effect of helium}


   \author{R. Soler \and R. Oliver \and J. L. Ballester}

   \offprints{R. Soler}

   \institute{Departament de F\'isica, Universitat de les Illes Balears,
              E-07122, Palma de Mallorca, Spain\\
              \email{[roberto.soler;ramon.oliver;joseluis.ballester]@uib.es}
             }


 	 \date{Received XXX / Accepted XXX}

  \abstract
   {Prominences are partially ionized, magnetized plasmas embedded in the solar corona. Damped oscillations and propagating waves are commonly observed. These oscillations have been interpreted in terms of magnetohydrodynamic (MHD) waves. Ion-neutral collisions and non-adiabatic effects (radiation losses and thermal conduction) have been proposed as damping mechanisms.}
   {We study the effect of the presence of helium on the time damping of non-adiabatic MHD waves in a plasma composed by electrons, protons, neutral hydrogen, neutral helium (\ion{He}{i}), and singly ionized helium (\ion{He}{ii}) in the single-fluid approximation.}
   {The dispersion relation of linear non-adiabatic MHD waves in a homogeneous, unbounded, and partially ion¡zed prominence medium is derived. The period and the damping time of Alfv\'en, slow, fast, and thermal waves are 
computed. A parametric study of the ratio of the damping time to the period with respect to the helium abundance is performed.}
   {The efficiency of ion-neutral collisions as well as thermal conduction is increased by the presence of helium. However, if realistic abundances of helium in prominences ($\sim 10\%$) are considered, this effect has a minor influence on the wave damping.}
   {The presence of helium can be safely neglected in studies of MHD waves in partially ionized prominence plasmas.}

   \keywords{Sun: oscillations --
                Sun: magnetic fields --
                Sun: corona --
		Sun: prominences
               }

	\titlerunning{MHD waves in a partially ionized prominence: Effect of helium}

   \maketitle
%


\section{Introduction}

Small-amplitude oscillations and propagating waves are commonly observed in both quiescent and active region prominences/filaments. They have been interpreted in terms of magnetohydrodynamic (MHD) eigenmodes of the magnetic structure and/or propagating MHD waves. The reader is referred to some recent reviews for more information about the observational and theoretical backgrounds \citep{oliverballester02,engvold04,ballester,banerjee,engvold08}

Prominence oscillations are known to be quickly damped, with damping times corresponding to a few oscillatory periods \citep[this topic has been reviewed by][]{oliver, mackay}. Several damping mechanisms of MHD waves have been proposed, non-adiabatic effects and ion-neutral collisions being the more extensively investigated. In order to understand in detail these effects, they have been studied in simple configurations such as unbounded and homogeneous media. \citet{carbonell04} investigated the time damping in a homogeneous prominence medium taking non-adiabatic effects (optically thin radiation losses and thermal conduction) into account. Later on, the spatial damping was studied by \citet{carbonell06} and the effect of a background mass flow was analyzed by \citet{carbonell09}. Subsequently, some works have extended these previous results by considering the presence of the coronal medium \citep{soler07,soler08,soler09NA}. The common conclusion of these investigations is that only slow and thermal waves are efficiently damped by non-adiabatic effects, while fast waves are very slightly damped and Alfv\'en waves are completely unaffected. 

On the other hand, the influence of partial ionization on the propagation and time damping of MHD waves has been also investigated in an unbounded medium. \citet{forteza07} followed the treatment by \citet{brag} and derived the full set of MHD equations along with the dispersion relation of linear waves in a partially ionized, single-fluid plasma \citep[see also][]{pinto}. The presence of electrons, protons, and neutral hydrogen atoms was taken into account, whereas helium and other species were not considered. In a subsequent work \citep{forteza08}, they extended their previous analysis by considering radiative losses and thermal conduction by electrons and neutrals. Their main results with respect to the fully ionized case \citep{carbonell04} were, first, that ion-neutral collisions (by means of the so-called Cowling's diffusion) can damp both Alfv\'en and fast waves but non-adiabatic effects remain only important for the damping of slow and thermal waves, and second, that there exist critical values of the wavenumber in which the real part of the frequency vanishes, so wave propagation is not possible for larger wavenumbers. Again, applications to a more complex cylindrical geometry have been also performed \citep{soler09IN,soler09INRA}

On the basis of these previous results, it seems clear that partial ionization plays a relevant role on wave propagation in prominences. Prominences are roughly composed by 90\% hydrogen and 10\% helium but, to date, all the investigations considered a pure hydrogen plasma. Therefore, the effect of the presence of helium on the propagation and damping of MHD waves is still unknown and is the motivation for the present work. Here, we consider an unbounded and homogeneous prominence medium permeated by a homogeneous magnetic field. The plasma is assumed to be partially ionized, electrons, protons, neutral hydrogen, neutral helium (\ion{He}{i}), and singly ionized helium (\ion{He}{ii}) being the species taken into account. Recent studies by \citet{labrosse} indicate that for central prominence temperatures, the ratio of the number densities of \ion{He}{ii} to \ion{He}{i} is around 10\%, whereas the presence of \ion{He}{iii} is negligible. This result allows us to neglect \ion{He}{iii} in this work. Extending the works by \citet{forteza07,forteza08}, the derivation of the basic MHD equations for a non-adiabatic, partially ionized, single-fluid plasma has been generalized by considering now five different species, allowing us to study how the presence of neutral and singly ionized helium affects their previous results.

This paper is organized as follows. The description of the equilibrium and the basic equations are given in Sect.~\ref{sec:equations}. The results are discussed in Sect.~\ref{subsec:results}. Finally, Sect.~\ref{sec:conclusions} contains the conclusion of this work.


 \section{Equilibrium and basic equations \label{sec:equations}} 
\begin{figure}[!htb]
\centering
\includegraphics[width=0.75\columnwidth]{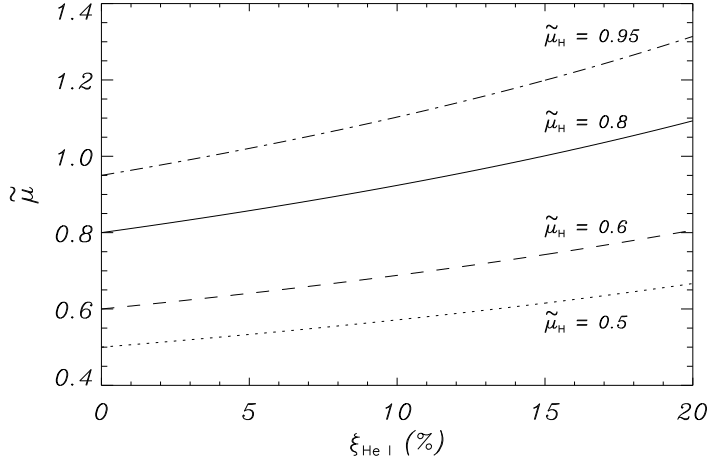}
\caption{Mean atomic weight, $\mutilde$, as a function of the relative neutral helium density, $\xi_\ion{He}{i}$, for $\mutilde_{\rm H} = 0.5$ (dotted line), $\mutilde_{\rm H} = 0.6$ (dashed line),  $\mutilde_{\rm H} = 0.8$ (solid line), and $\mutilde_{\rm H} = 0.95$ (dash-dotted line). In all cases, $\delta_{\rm He} = 0.1$.\label{fig:mu}}
\end{figure}

 Our equilibrium configuration is a homogeneous and unbounded partially ionized plasma composed by electrons, protons, neutral hydrogen, neutral helium, and singly ionized helium. Hereafter, subscripts e, p, H, \ion{He}{i}, and \ion{He}{ii} explicitly denote these species, respectively. The magnetic field is also homogeneous and orientated along the $x$-direction, $\vec{B}_0=B_0 \hat{e}_x$, with $B_0 = 5$~G. We adopt the single-fluid approximation. Following \citet{forteza07,forteza08} and neglecting the electron contribution, we define the center of mass velocity, $\vec{v}$, as follows,
\begin{equation}
 \vec{v} \approx \xi_{\rm p} \vec{v}_{\rm p} +  \xi_{\rm H} \vec{v}_{\rm H} + \xi_{\ion{He}{i}} \vec{v}_{\ion{He}{i}} + \xi_{\ion{He}{ii}} \vec{v}_{\ion{He}{ii}},
\end{equation}
with $\xi_\alpha$ the relative density of species $\alpha$, and $\vec{v}_\alpha$ the corresponding species velocity. Equivalently, the equilibrium total density, $\rho_0$, and gas pressure, $p_0$, are,
\begin{equation}
 \rho_0 \approx \rho_{\rm p} + \rho_{\rm H} + \rho_{\ion{He}{i}} + \rho_{\ion{He}{ii}}, 
\end{equation}
\begin{equation}
p_0 = 2 \left( p_{\rm p} + p_{\ion{He}{ii}} \right) + p_{\rm H} + p_{\ion{He}{i}}.
\end{equation}
Since $\rho_\alpha = \xi_\alpha \rho_0$, we get the relation $\xi_{\rm p} + \xi_{\rm H} + \xi_\ion{He}{i} + \xi_{\ion{He}{ii}} \approx 1$. We assume a strong thermal coupling between species, so all the species have the same equilibrium temperature $T_0$. Then, the three equilibrium quantities are related as follows,
\begin{equation}
 p_0 = \rho_0\frac{R}{\mutilde} T_0,
\end{equation}
where $R$ is the ideal gas constant and $\mutilde$ is the mean atomic weight, 
\begin{equation}
 \mutilde = \frac{1}{2 \xi_{\rm p} + \xi_{\rm H} + \frac{1}{4}\xi_\ion{He}{i} + \frac{1}{2}\xi_{\ion{He}{ii}}}.
\end{equation}
With the help of some definitions, we can express $\mutilde$ in a more convenient form,
\begin{equation}
 \mutilde = \frac{\mutilde_{\rm H}}{1- \left[\left( 1 + \delta_{\rm He} \right)+ \left( 1 + 2\delta_{\rm He} \right)\frac{1}{4}\mutilde_{\rm H}\right] \xi_\ion{He}{i}}, \label{eq:mut}
\end{equation}
with 
\begin{equation}
 \mutilde_{\rm H} = \frac{\xi_{\rm p} + \xi_{\rm H}}{2\xi_{\rm p} + \xi_{\rm H}}, \qquad \delta_{\rm He} = \frac{\xi_{\ion{He}{ii}}}{\xi_\ion{He}{i}}.
\end{equation}
The quantity $\mutilde_{\rm H}$ is equivalent to the mean atomic weight of a pure hydrogen plasma defined in Eq.~(3) of \citet{forteza07}, and  ranges between $\mutilde_{\rm H} = 0.5$ for a fully ionized hydrogen plasma and $\mutilde_{\rm H} = 1$ for a fully neutral hydrogen gas. On the other hand, $\delta_{\rm He}$ indicates the helium ionization degree. A realistic value of this parameter is $\delta_{\rm He} = 0.1$ according to the results of \citet{labrosse}. From Eq.~(\ref{eq:mut}) one can see that $\mutilde > \mutilde_{\rm H}$ due to the presence of helium. In the absence of helium, $\xi_\ion{He}{i} =\xi_{\ion{He}{ii}}=0$ so $\mutilde = \mutilde_{\rm H}$. Figure~\ref{fig:mu} displays the dependence of $\mutilde$ on $\xi_\ion{He}{i}$ for several values of $\mutilde_{\rm H}$. 

The details of the derivation of the basic governing equations for a non-adiabatic, partially ionized, one-fluid plasma can be followed in, e.g., \citet{brag,forteza07,forteza08,pinto}. Here, we follow the same procedure but generalize the analysis of \citet{forteza07} by including additional species. In brief, the separate governing equations for the five species are added and a generalized Ohm's law is obtained. These basic equations correspond to Eqs.~(1)--(6) of \citet{forteza08}, which are formally identical in our case. A key step in the present derivation is to compute the density current, $\vec{j}$, as
\begin{equation}
 \vec{j} = e \left( n_{\rm p} \vec{v}_{\rm p} + n_\ion{He}{ii} \vec{v}_\ion{He}{ii} - n_{\rm e} \vec{v}_{\rm e} \right),
\end{equation}
along with the condition $n_{\rm e} = n_{\rm p}  + n_\ion{He}{ii}$, where $n_{\rm e}$, $n_{\rm p}$, and $n_\ion{He}{ii}$ are the electron, proton, and \ion{He}{ii} number densities, respectively, and $e$ is the electron charge. The resulting induction equation (see Eq.~[14] of \citealt{forteza07}) contains several diffusion terms whose coefficients depend on the collisional frequencies between species. The physical meaning of these nonideal terms is explained in detail in \citet{pinto}. In particular, ion-neutral collisions are responsible for the so-called Cowling's diffusion, which is much more efficient than Ohm's diffusion in a partially ionized plasma. However, some terms are not relevant for our present application. Hall's effect is negligible in prominence conditions \citep{soler09INRA}, and the so-called ``Biermann's battery'' term is identically zero in a homogeneous medium. For this reason, our final form of the induction equation (Eq.~[21] of \citealt{forteza07}) only contains the terms corresponding to Ohm's and ambipolar (Cowling's) diffusion, along with the diamagnetic current term.

Ohm's, $\eta$, and Cowling's, $\eta_{\rm C}$, coefficients of magnetic diffusion can be expressed in terms of their corresponding conductivities,
\begin{equation}
  \eta = \frac{1}{\mu \sigma}, \qquad  \eta_{\rm C} = \frac{1}{\mu \sigma_{\rm C}},
\end{equation}
with $\mu = 4\pi \times 10^{-7}$~N~A$^{-2}$. So, Ohm's and Cowling's conductivities, as well as the diamagnetic current coefficient, $\Xi$, which applies in our case when helium is included are:
\begin{equation}
 \sigma = \frac{e^2 n_{\rm e}^2}{\left(\alpha_{\rm e} - \alpha_{\rm en}^2 / \alpha_{\rm n}\right)}, \qquad  \sigma_{\rm C} = \frac{\sigma}{1 + \frac{B_0^2 \left(\xi_{\rm H} + \xi_\ion{He}{i}  \right)^2 }{ \alpha_{\rm n}}\sigma }
\end{equation}
\begin{equation}
 \Xi = \frac{\left( \xi_{\rm H} + \xi_\ion{He}{i} \right)}{\mutilde\, \alpha_{\rm n}} \left( \xi_{\rm p} \xi_{\rm H} - \frac{1}{2}\xi_{\rm H} \xi_{\ion{He}{ii}} + \frac{7}{4}\xi_{\rm p}\xi_\ion{He}{i} + \frac{1}{4}\xi_\ion{He}{i} \xi_{\ion{He}{ii}} \right).
\end{equation}
In addition, $\alpha_{\rm e}$, $\alpha_{\rm en}$, and $\alpha_{\rm n}$ are the electron, electron-neutral, and neutral friction coefficients, respectively, whose expressions depend on the sum of the friction coefficients between particular species,
\begin{equation}
 \alpha_{\rm en} =  \alpha_{\rm eH} + \alpha_{\rm e\ion{He}{i}},
\end{equation}
\begin{equation}
 \alpha_{\rm e} =  \alpha_{\rm ep} + \alpha_{\rm eH} + \alpha_{\rm e\ion{He}{i}} + \alpha_{\rm e\ion{He}{ii}},
\end{equation}
\begin{equation}
 \alpha_{\rm n} =  \alpha_{\rm eH} + \alpha_{\rm e\ion{He}{i}} +\alpha_{\rm pH} + \alpha_{\rm p\ion{He}{i}} + \alpha_{\rm \ion{He}{ii}H} + \alpha_{\rm \ion{He}{ii}\ion{He}{i}}.
\end{equation}
Each particular friction coefficient, $\alpha_{\beta \beta'}$, is computed as
\begin{equation}
 \alpha_{\beta \beta'} = n_\beta m_{\beta \beta'} \nu_{\beta \beta'},
\end{equation}
with $n_\beta$ the number density of the species $\beta$, $\nu_{\beta \beta'}$ the collisional frequency between species $\beta$ and $\beta'$, and
\begin{equation}
 m_{\beta \beta'} = \frac{m_\beta m_{\beta'}}{m_\beta + m_{\beta'}},
\end{equation}
with $m_\beta$ the mass particle of the species $\beta$. As given by \citet{depontieu}, see also \citet{soler09IN}, the collisional frequencies between electrons and protons or \ion{He}{ii} are
\begin{equation}
 \nu_{\rm e i} = 3.7 \times 10^{-6} \frac{n_{\rm i} \ln\Lambda}{T_0^{3/2}}, 
\end{equation}
with ${\rm i} = {\rm p}$ or $\ion{He}{ii}$ and $\ln\Lambda$ the Coulomb logarithm, while the collisional frequency between a charged species, $\rm q= {\rm e}$, ${\rm p}$, or $\ion{He}{ii}$, and a neutral species, $\rm n={\rm H}$ or $\ion{He}{i}$, is 
\begin{equation}
 \nu_{{\rm q n}} = n_{\rm n} \sqrt{\frac{8 k_{\rm B} T_0}{\pi m_{{\rm q n}}}}\Sigma_{{\rm q n}},
\end{equation}
with $k_{\rm B}$ the Boltzmann's constant, and $\Sigma_{{\rm q n}}$ the collisional cross-section. Here, we consider the values $\Sigma_{{\rm e n}} = 10^{-19}$~m$^2$, and $\Sigma_{{\rm p n}} =\Sigma_{{\rm \ion{He}{ii} n}}= 5 \times 10^{-19}$~m$^2$.

\begin{figure*}[!htp]
\centering
\includegraphics[width=0.66\columnwidth]{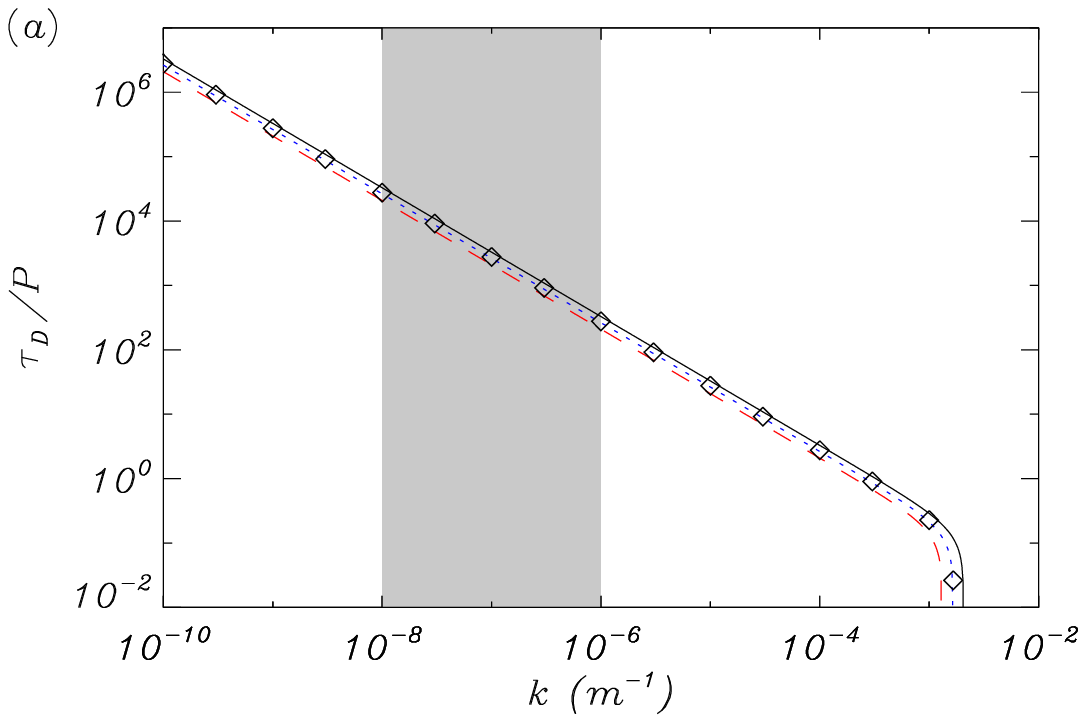}
\includegraphics[width=0.66\columnwidth]{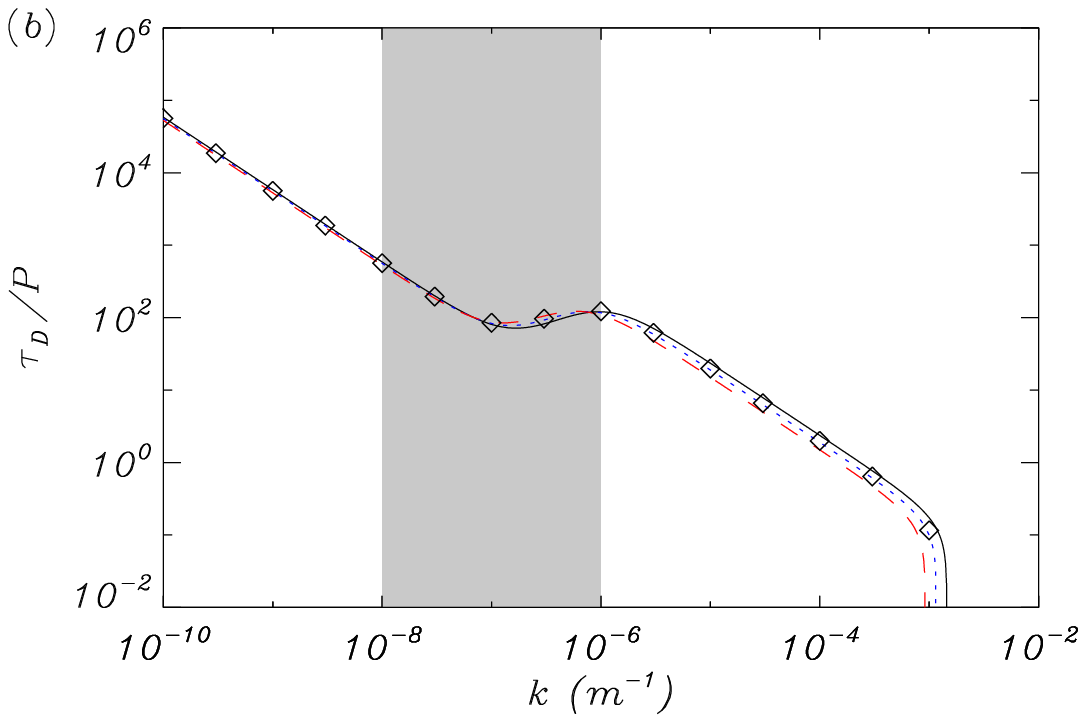}
\includegraphics[width=0.66\columnwidth]{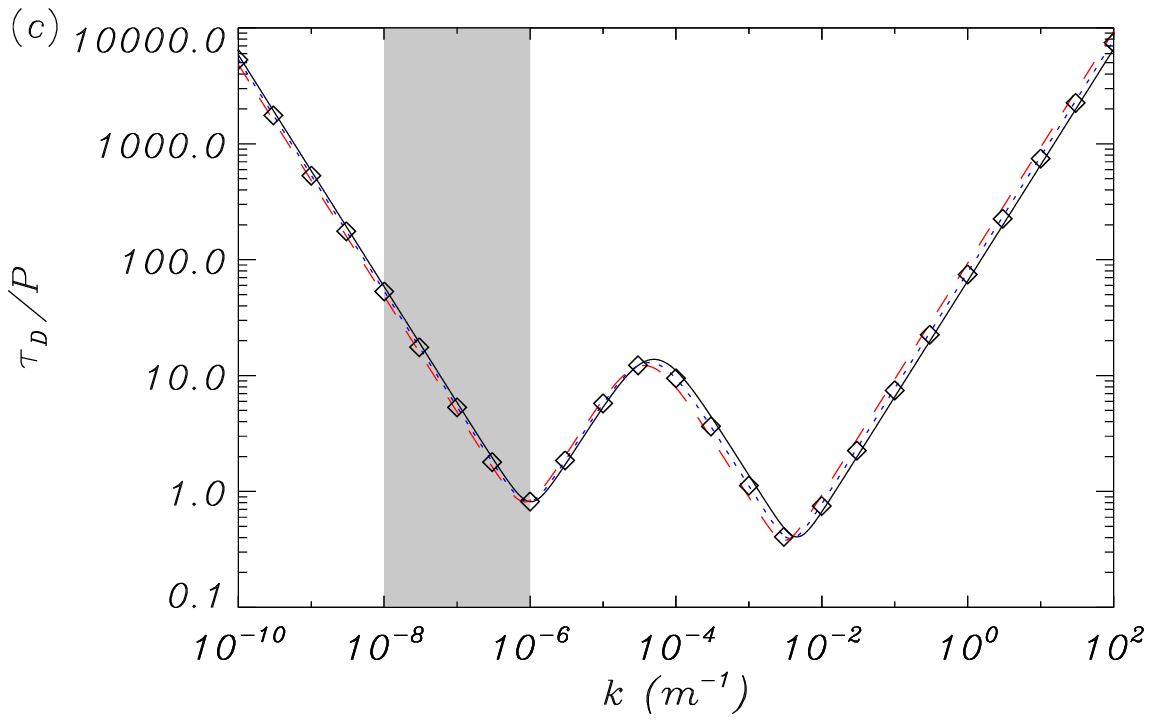}
\caption{Ratio of the damping time to the period, $\tdp$, versus the wavenumber, $k$, corresponding to the ($a$) Alfv\'en wave, ($b$) fast wave, and ($c$) slow wave for $\theta = \pi/4$, $\mutilde_{\rm H} = 0.8$, and $\delta_{\rm He} = 0.1$. The different linestyles represent $\xi_\ion{He}{i} = 0\%$ (solid line), $\xi_\ion{He}{i} = 10\%$ (dotted line), and $\xi_\ion{He}{i} = 20\%$ (dashed line). The results for $\xi_\ion{He}{i} = 10\%$ and $\delta_{\rm He} = 0.5$ are plotted by means of symbols for comparison. The shaded regions correspond to the range of typically observed wavelengths of prominence oscillations. \label{fig:mhdwaves}}
\end{figure*}


 On the other hand, the thermal conductivity due to neutrals (Eq.~[16] of \citealt{forteza08}) has to include now the helium contribution. According to \citet{parker}, a corrected expression for the conductivity of neutrals in MKS units is 
\begin{equation}
 \kappa_{\rm n}  = \kappa_{\rm H} + \kappa_\ion{He}{i} =  \left( 2.44 \times 10^{-2} \xi_{\rm H} +   3.18 \times 10^{-2} \xi_\ion{He}{i} \right) T_0^{1/2}.     \label{eq:cond}
\end{equation}
Finally, we assume an optically thin radiation \citep{hildner} to represent the hydrogen radiative losses. According to \citet[see their Fig.~3]{coxtucker}, the radiative losses by helium are several orders of magnitude smaller than those of hydrogen for typical prominence temperatures ($\sim 10^4$~K) and, therefore, irrelevant for the present investigation.

Hereafter, our analysis follows that of \citet{forteza08}. We linearize the basic equations and assume small perturbations proportional to $\exp \left( i \omega t + i k_x x + i k_z z \right)$. Then the resulting equations (Eq.~[18]--[27] of \citealt{forteza08}) are combined and finally two different, uncoupled dispersion relations, one for Alfv\'en waves (their Eq.~[28]) and another for magnetoacoustic and thermal waves (their Eq.~[30]) are obtained. Note that although our definitions of $\eta$, $\eta_{\rm C}$, $\Xi$, and $\kappa_{\rm n}$ contain the effect of helium, the resulting dispersion relations are formally identical to those of \citet{forteza08}. For the sake of simplicity, we do not write again these expressions here and refer the reader to \citet{forteza08}. The dispersion relations are numerically solved for real values of the wavenumber modulus, $k = \sqrt{k_x^2 + k_z^2}$, and the angle $\theta$ between $\vec{B}_0$ and $\vec{k}$. A complex frequency, $\omega = \omega_{\rm R} + i \omega_{\rm I}$, is obtained. The period, $P$, and damping time, $\td$, are related to the real and imaginary parts of the frequency as follows,
\begin{equation}
 P = \frac{2 \pi}{\omega_{\rm R}}, \qquad \tau_{\rm D} = \frac{1}{\omega_{\rm I}}.
\end{equation}

%
\section{Results}
\label{subsec:results}

In the following computations, we consider typical prominence conditions, $\rho_0 = 5 \times 10^{-11}$~kg~m$^{-3}$ and $T_0=8000$~K. Quantities $\mutilde_{\rm H}$, $\xi_\ion{He}{i}$, and $\delta_{\rm He}$ are considered free parameters. We focus our attention on the effect of the relative neutral helium density, $\xi_\ion{He}{i}$, on the ratio $\tdp$. 

\subsection{Free propagation in an unbounded medium}

First, we assume $\theta = \pi/4$. Figure~\ref{fig:mhdwaves} displays $\tdp$ as a function of $k$ for the Alfv\'en, fast, and slow waves. The results corresponding to several helium abundances are compared for hydrogen and helium ionization degrees of $\mutilde_{\rm H} = 0.8$ and $\delta_{\rm He}=0.1$, respectively. We see that even in the case of the largest quantity of helium considered ($\xi_\ion{He}{i} = 20\%$), the presence of helium has a minor effect on the results. In the case of Alfv\'en and fast waves (Fig.~\ref{fig:mhdwaves}a,b), their critical wavenumber (i.e., the value of $k$ which causes the real part of the frequency to vanish) is shifted toward slightly smaller values. So, the larger $\xi_\ion{He}{i}$, the smaller $k_{\rm c}^{\rm a}$. This result can be understood by considering that the Alfv\'en wave critical wavenumber, $k_{\rm c}^{\rm a}$, given by Eq.~(38) of \citet{forteza08} is,
\begin{equation}
 k_{\rm c}^{\rm a} = \frac{2 \va}{\left( \eta_{\rm C} + \eta \tan^2 \theta \right) \cos \theta}, \label{eq:crit}
\end{equation}
with $\va = B_0 / \sqrt{\mu \rho_0}$ the Alfv\'en speed. Equation~(\ref{eq:crit}) is also approximately valid for the fast wave critical wavenumber. Then, we see that $k_{\rm c}^{\rm a}$ is inversely proportional to Cowling's diffusivity, $\eta_{\rm C}$. Since $\eta_{\rm C}$ is larger in the presence of helium  than in the pure hydrogen case due to additional collisions of neutral and singly ionized helium species,  $k_{\rm c}^{\rm a}$ is therefore smaller. Turning our attention to the slow wave (Fig.~\ref{fig:mhdwaves}c), we see that the maximum and the right-hand side minimum of $\tdp$ are also slightly shifted toward smaller values of $k$. Results from \citet{carbonell04} and \citet{forteza08} indicate that thermal conduction is responsible for these maximum and minimum of $\tdp$. Thus, the additional contribution of neutral helium atoms to thermal conduction (Eq.~\ref{eq:cond}) causes this displacement of the curve of $\tdp$. As for Alfv\'en and fast waves, this effect is of minor importance. For comparison, equivalent results with $\xi_\ion{He}{i} = 10\%$ and $\delta_{\rm He} = 0.5$ are plotted by means of symbols in Fig.~\ref{fig:mhdwaves}. We see that for realistic values of $\delta_{\rm He}$, its role is almost irrelevant, meaning that the presence of \ion{He}{ii} can be neglected. It is worth mentioning that we have repeated these calculations for other values of $\mutilde_{\rm H}$ and similar results have been obtained.

\begin{figure}[!htb]
\centering
\includegraphics[width=0.75\columnwidth]{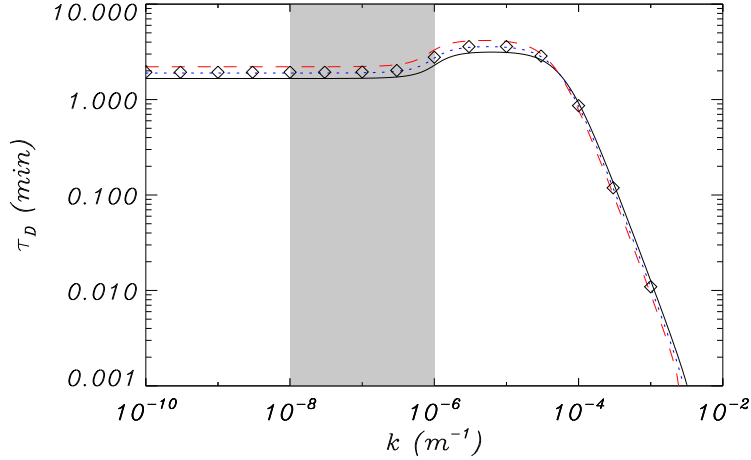}\\
\caption{Damping time, $\td$, of the thermal wave versus the wavenumber, $k$, with $\theta = \pi/4$. The different linestyles represent: $\xi_\ion{He}{i} = 0\%$ (solid line), $\xi_\ion{He}{i} = 10\%$ (dotted line), and $\xi_\ion{He}{i} = 20\%$ (dashed line). In all computations, $\mutilde_{\rm H} = 0.8$ and $\delta_{\rm He} = 0.1$. The result for $\xi_\ion{He}{i} = 10\%$ and $\delta_{\rm He} = 0.5$ is plotted by means of symbols for comparison. \label{fig:therm}}
\end{figure}

Next, we study the thermal mode. Since it is a purely damped, non-propagating disturbance ($\omega_{\rm R} = 0$), we only plot  the damping time, $\td$, as a function of $k$ for $\mutilde_{\rm H} = 0.8$ and $\delta_{\rm He}=0.1$ (Fig.~\ref{fig:therm}). We can see that the effect of helium is different in two ranges of $k$. For $k \gtrsim 10^{-4}$~m$^{-1}$, thermal conduction is the dominant damping mechanism. So, the larger the amount of helium, the smaller $\td$ because of the enhanced thermal conduction by neutral helium atoms. On the other hand, radiative losses are more relevant for $k \lesssim 10^{-4}$~m$^{-1}$. In this region, the thermal mode damping time grows as the helium abundance increases.  Since these variations of the damping time are very small, we have to conclude again that the damping time obtained in the absence of helium does not significantly change when helium is taken into account. Computations with other values of $\mutilde_{\rm H}$ and $\delta_{\rm He}$ do not modify this statement.


\subsection{Constrained propagation by a waveguide}

\begin{figure*}[!htp]
\centering
\includegraphics[width=0.66\columnwidth]{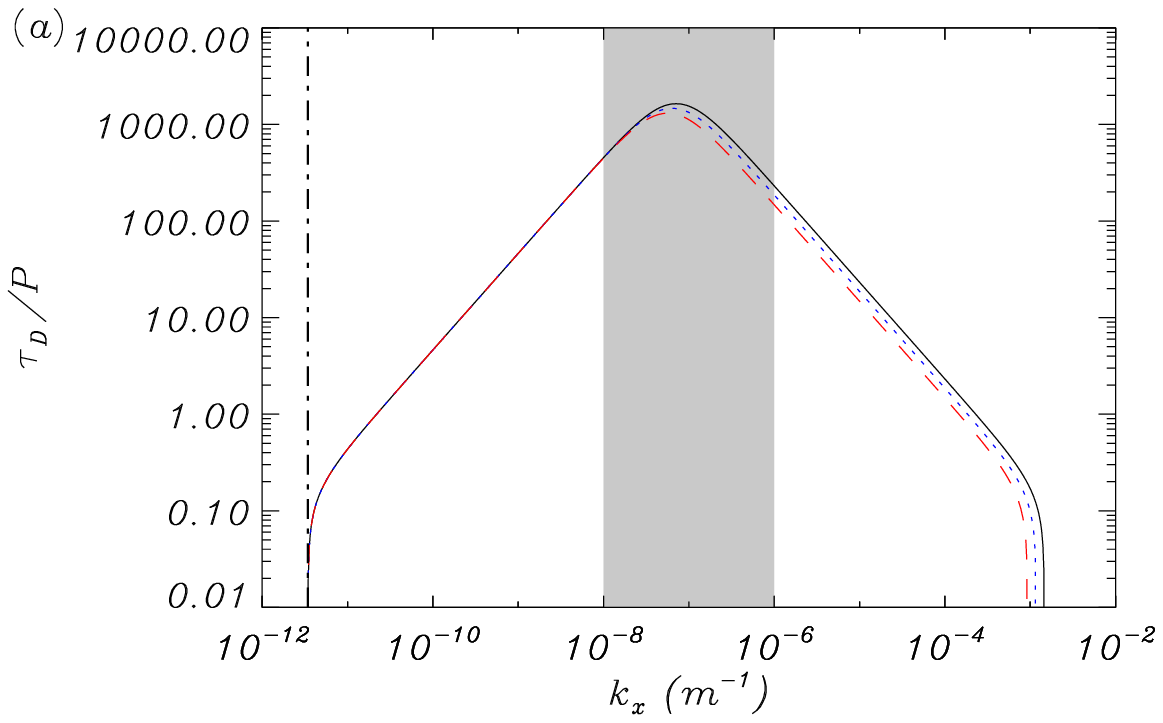}
\includegraphics[width=0.66\columnwidth]{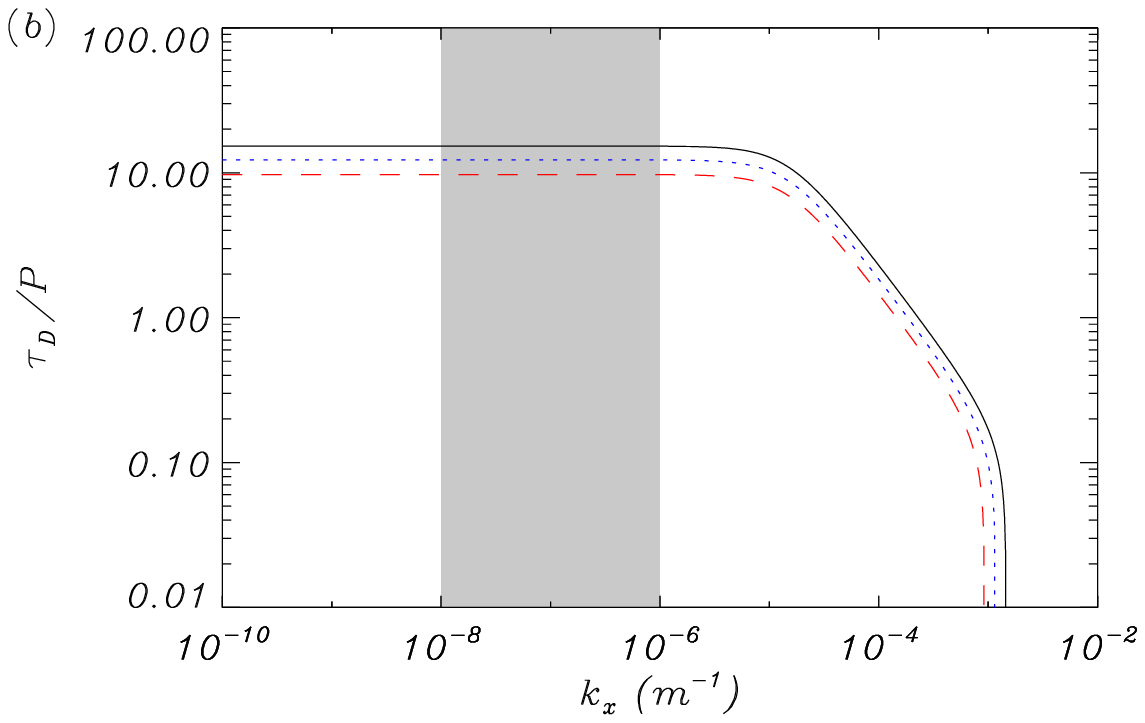}
\includegraphics[width=0.66\columnwidth]{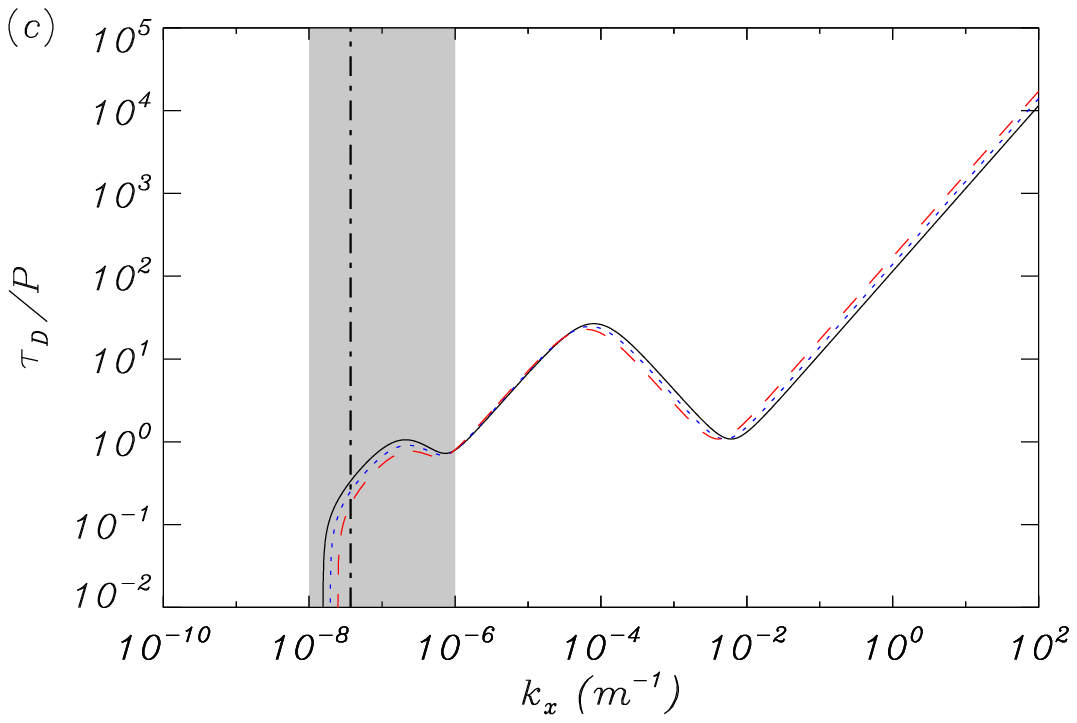}
\caption{Ratio of the damping time to the period, $\tdp$, versus the wavenumber component parallel to magnetic field lines, $k_x$, corresponding to the ($a$) Alfv\'en wave, ($b$) fast wave, and ($c$) slow wave for $k_z L = \pi/2$, with $L=10^5$~m, $\mutilde_{\rm H} = 0.8$, and $\delta_{\rm He} = 0.1$. The different linestyles represent $\xi_\ion{He}{i} = 0$ (solid line), $\xi_\ion{He}{i} = 10\%$ (dotted line), and $\xi_\ion{He}{i} = 20\%$ (dashed line). The vertical dot-dashed lines in ($a$) and ($c$) correspond to the approximated critical wavenumbers given by Eqs.~(\ref{eq:kcrtalf}) and (\ref{eq:kcrtslow}), respectively, for $\xi_\ion{He}{i} = 10\%$. \label{fig:mhdwaveskz}}
\end{figure*}

We can estimate the effect of a magnetic structure, say a slab or a cylinder, which would act as a waveguide. To do so, we set the wavenumber component in the perpendicular direction to magnetic field lines to a fixed value, $k_z L = \pi/2$, with $L$ a typical length-scale in the perpendicular direction. Since high-resolution observations of filaments \citep[see, e.g.,][]{lin07,lin09} show fine-structures (threads) with a typical width of $\sim 100$~km, we select $L = 10^{5}$~m as our perpendicular length-scale. Therefore, the propagation angle $\theta$ depends now on $k_x$,
\begin{equation}
 \theta = \arctan \left( \frac{\pi/2}{k_x L} \right).
\end{equation}
Figure~\ref{fig:mhdwaveskz} displays the results for the Alfv\'en, fast, and slow waves. We see that the behavior of the three solutions is substantially different from that of the free propagation case. The Alfv\'en mode now possesses an additional critical wavenumber for small values of $k_x$, namely $k_x^{\rm c -}$, which is independent of the ionization degree and the helium abundance. It can be approximated as (see details in \citealt{soler09IN}, Eq.~[38]),
\begin{equation}
 k_x^{\rm c -} \approx \frac{\eta}{2 \va} k_z^2= \frac{\eta \pi^2}{8 \va L^2}. \label{eq:kcrtalf}
\end{equation}
On the other hand, the fast wave is now more attenuated in the relevant range of wavenumbers than in the free propagation case, whereas the slow wave also has a new critical wavenumber, namely $k_x^{\rm c s}$ which falls within the relevant range. An expression for the slow mode critical wavenumber is also provided by Eq.~(48) of \citet{soler09IN}, which in our present notation is,
\begin{equation}
 k_x^{\rm c s} \approx \frac{\cs \eta_{\rm C}  }{2 \va^2}k_z^2  =  \frac{\cs \eta_{\rm C} \pi^2 }{8 \va^2 L^2}, \label{eq:kcrtslow}
\end{equation}
where $\cs = \sqrt{\gamma p_0 / \rho_0}$ is the sound speed, with $\gamma = 5/3$ the adiabatic index. The slow mode critical wavenumber is shifted toward larger values as the helium abundance increases. Note that there is no additional critical wavenumber for the fast wave. These approximated critical wavenumbers are indicated by means of vertical lines in Fig.~\ref{fig:mhdwaveskz}. We see an excellent agreement in the case of the Alfv\'en mode critical wavenumber (Eq.~[\ref{eq:kcrtalf}]). For the slow wave, the approximated value (Eq.~[\ref{eq:kcrtslow}]) is slightly larger than that numerically obtained.

Finally, we have also computed the results in the case of the guided thermal disturbance. We find that the thermal mode behavior is the same in the waveguide case and in the free propagation case. Hence, this mode is not affected by the variation of the propagation angle and no further comments are needed.

\section{Conclusion}
\label{sec:conclusions}

In this work, we have studied the effect of helium (\ion{He}{i} and \ion{He}{ii}) on the time damping of thermal and MHD waves in a partially ionized prominence plasma. This is an extension of previous investigations by \citet{forteza07,forteza08} in which helium was not taken into account. We conclude that, although the presence of neutral helium increases the efficiency of both ion-neutral collisions and thermal conduction, its effect is not important for realistic helium abundances in prominences. In addition, due to the very small \ion{He}{ii} abundance for central prominence temperatures, its presence is irrelevant to the wave behavior. This conclusion applies both to the free propagation case and the constrained propagation by a waveguide case. Although the role of \ion{He}{ii} (or even \ion{He}{iii}) could be larger for typical prominence-corona transition region temperatures, the present result allows future studies of MHD waves and oscillations in prominences to neglect the presence of helium.

\begin{acknowledgements}
      We thank N. Labrosse for giving useful information about the helium ionization degree in prominences and I. Arregui for some useful comments. The authors acknowledge the financial support received from the Spanish MICINN, FEDER funds, and the Conselleria d'Economia, Hisenda i Innovaci\'o of the CAIB under Grants No. AYA2006-07637 and PCTIB-2005GC3-03. RS thanks the CAIB for a fellowship.
\end{acknowledgements}

\bibliographystyle{aa} 

\begin{thebibliography}{}

  \bibitem[Ballester(2006)]{ballester} Ballester, J. L. 2006, Phil. Trans. R. Soc. A, 364, 405 
  \bibitem[Banerjee et al.(2007)]{banerjee} Banerjee, D., Erd\'elyi, R., Oliver R., \& O'Shea, E. 2007, \solphys, 246, 3

\bibitem[Braginskii(1965)]{brag} Braginskii, S. I. 1965, Rev. Plasma Phys., 1, 205

   \bibitem[Carbonell et al.(2004)]{carbonell04} Carbonell, M., Oliver, R., \& Ballester, J. L. 2004, \aap, 415, 739 
%
   \bibitem[Carbonell et al.(2006)]{carbonell06} Carbonell, M., Terradas, J., Oliver, R., \& Ballester, J. L. 2006, \aap, 460, 573 	
%

\bibitem[Carbonell et al.(2009)]{carbonell09} Carbonell, M., Oliver, R., \& Ballester, J. L. 2009, NewA, 14, 277 
%

\bibitem[Cox \& Tucker(1969)]{coxtucker} Cox, D. P., \& Tucker, W. H. 1969, \apj, 157, 1157

\bibitem[De Pontieu et al.(2001)]{depontieu} De Pontieu, B., Martens, P. C. H., \& Hudson, H. S. 2001, \apj, 558, 859

  \bibitem[Engvold(2004)]{engvold04} Engvold, O. 2004, Proc. IAU Collq. on Multiwavelength investigations of solar activity (eds. A. V. Stepanov, E. E. Benevolenskaya \& A. G. Kosovichev), 187
%
%
\bibitem[Engvold(2008)]{engvold08} Engvold, O. 2008, in IAU Symp. 247, Waves \& Oscillations in the Solar Atmosphere: Heating and Magneto-Seismology, ed. R. Erd\'elyi \& C. A. Mendoza-Brice\~no (Cambridge: Cambridge Univ. Press), 152

   \bibitem[Forteza et al.(2007)]{forteza07} Forteza, P., Oliver, R., Ballester, J. L. \& Khodachenko, M. L. 2007, \aap, 461, 731

 \bibitem[Forteza et al.(2008)]{forteza08} Forteza, P., Oliver, R., \& Ballester, J. L. 2008, \aap, 492, 223
%
%

\bibitem[Gouttebroze \& Labrosse(2009)]{labrosse} Gouttebroze, P., \& Labrosse, N. 2009, \aap, 503, 663

   \bibitem[Hildner(1974)]{hildner} Hildner, E. 1974, \solphys, 35, 123
%

 \bibitem[Lin et al.(2007)]{lin07} Lin, Y., Engvold, O., Rouppe van der Voort, L. H. M., \& van Noort, M.  2007, \solphys, 246, 65

\bibitem[Lin et al.(2009)]{lin09} Lin, Y., Soler, R., Engvold, O., Ballester, J. L., Langangen, \O., Oliver, R., \& Rouppe van der Voort, L. H. M. 2009, \apj, 704, 870


  \bibitem[Mackay et al.(2009)]{mackay} Mackay, D. H., Karpen, J. T., Ballester, J. L., Schmieder, B., \& Aulanier, G. 2009, Space Sci Rev, submitted
%
%
   \bibitem[Oliver \& Ballester(2002)]{oliverballester02} Oliver, R. \& Ballester, J. L. 2002, \solphys, 206, 45
 \bibitem[Oliver(2009)]{oliver} Oliver, R. 2009, Space Sci Rev, in press

   \bibitem[Parker(1953)]{parker} Parker, E. N. 1953, \apj, 117, 431

\bibitem[Pinto et al.(2008)]{pinto} Pinto, C., Galli, D., \& Bacciotti, F. 2008, \aap, 484, 1

%
\bibitem[Soler et al.(2007)]{soler07} Soler, R., Oliver, R., \& Ballester, J. L. 2007,  \aap, 471, 1023 
  \bibitem[Soler et al.(2008)]{soler08} Soler, R., Oliver, R., \& Ballester, J. L. 2008,  \apj, 684, 725 
  \bibitem[Soler et al.(2009a)]{soler09NA} Soler, R., Oliver, R., \& Ballester, J. L. 2009a, NewA, 14, 238 
 \bibitem[Soler et al.(2009b)]{soler09IN} Soler, R., Oliver, R., \& Ballester, J. L. 2009b,  \apj, 699, 1553 
\bibitem[Soler et al.(2009c)]{soler09INRA} Soler, R., Oliver, R., \& Ballester, J. L. 2009c,  \apj, submitted
%

\end{thebibliography}

%

\end{document}